# Superinjection of holes in homojunction diodes based on wide-bandgap semiconductors

Igor A. Khramtsov and Dmitry Yu. Fedyanin*

*Laboratory of Nanooptics and Plasmonics, Moscow Institute of Physics and Technology, 141700 Dolgoprudny, Russian Federation*
\* Correspondence: dmitry.fedyanin@phystech.edu

## Abstract

Electrically driven light sources are essential in a wide range of applications, from indication and display technologies to high-speed data communication and quantum information processing. Wide-bandgap semiconductors promise to advance solid-state lighting by delivering novel light sources. However, electrical pumping of these devices is still a challenging problem. Many wide-bandgap semiconductor materials, such as SiC, GaN, AlN, ZnS, and $Ga_2O_3$, can be easily doped n-type, but their efficient p-type doping is extremely difficult. The lack of holes due to the high activation energy of acceptors greatly limits the performance and practical applicability of wide-bandgap semiconductor devices. Here, we study a novel effect which allows homojunction semiconductors devices, such as p-i-n diodes, to operate well above the limit imposed by doping of the p-type material. Using a rigorous numerical approach, we show that the density of injected holes can exceed the density of holes in the p-type injection layer by up to three orders of magnitude, which gives the possibility to significantly overcome the doping problem. We present a clear physical explanation of this unexpected feature of wide-bandgap semiconductor p-i-n diodes and closely examine it in 4H-SiC, 3C-SiC, AlN and ZnS structures. The predicted effect can be exploited to develop bright light emitting devices, especially electrically driven non-classical light sources based on color centers in SiC, AlN, ZnO and other wide-bandgap semiconductors.

## Introduction

The possibility to create a high density of non-equilibrium charge carriers in the active region of semiconductor optoelectronic devices is essential for a wide range of optoelectronic applications, from LEDs [1] and injection lasers [2,3] to electro-optic modulators [4,5] and recently emerged single-photon sources [6–8]. The higher the carrier density, the better the performance is achieved. In most devices, the excess carriers are injected to the active region of the device from the heavily doped electron-rich n-type layers and hole-rich p-type layers, respectively [9]. However, wide-bandgap semiconductors are quite unique materials, which are at the interface between conventional semiconductors and insulators. They can demonstrate both the n-type and p-type conductivity but typically suffer from the lack of either electrons or holes due to the extremely high activation energies of dopants and non-zero compensation of donors (acceptors) by acceptor-type (donor-type) impurities [10–14]. Due to this problem, the density of free carriers in the n-type or p-type injection layer can be many orders of magnitude lower than the donor or acceptor concentration, respectively. The failure to produce enough free carriers is very often the only reason why the performance of light emitting devices based on wide-bandgap semiconductors is significantly lower than that of based on conventional semiconductor materials, such as gallium arsenide.

The traditional solution to this problem is to use the superinjection effect in double heterostructures [15–17]. As illustrated in Figure 1a, high potential barriers at the heterojunctions prevent electrons and holes from escaping from the central active region of the structure. Thus, under forward bias, all carriers injected from the n-type and p-type layers are confined to the central region of the double heterostructures. The densities of injected electrons and holes can be orders of magnitude higher than in the n-type and p-type injection layers (Figure 1b) [18], which is referred to as superinjection [15,16]. However, to exploit the superinjection effect in the particular semiconductor, one needs an auxiliary semiconductor material, which has a larger bandgap properly aligned with respect to the conduction and valence band edges of the considered semiconductor (Figure 1a), and with a lattice constant equal or close to that of the considered semiconductor. However, such an auxiliary material does not exist for many wide-bandgap semiconductors (see Figure 1.9 in [19]). Therefore, only homojunction structures can often be used.

For decades it was believed that the superinjection effect was a unique feature of semiconductor heterostructures, while the maximum density of injected electrons and holes in homojunction structures could not exceed the free carrier densities in the n-type and p-type injection layers, respectively [20]. However, recently, it was discovered that the superinjection of electrons is possible in homojunction diamond p-i-n diodes [14,21,22]: at a high forward bias voltage, a potential well for electrons is formed in the i-region of the diamond p-i-n diode in the vicinity of the p-i junction (see Figure 1c). Free electrons injected from the n-type layer are accumulated in this potential well so that their density can exceed the electron density in the n-type injection layer by orders of magnitude [22] (Figure 1d).

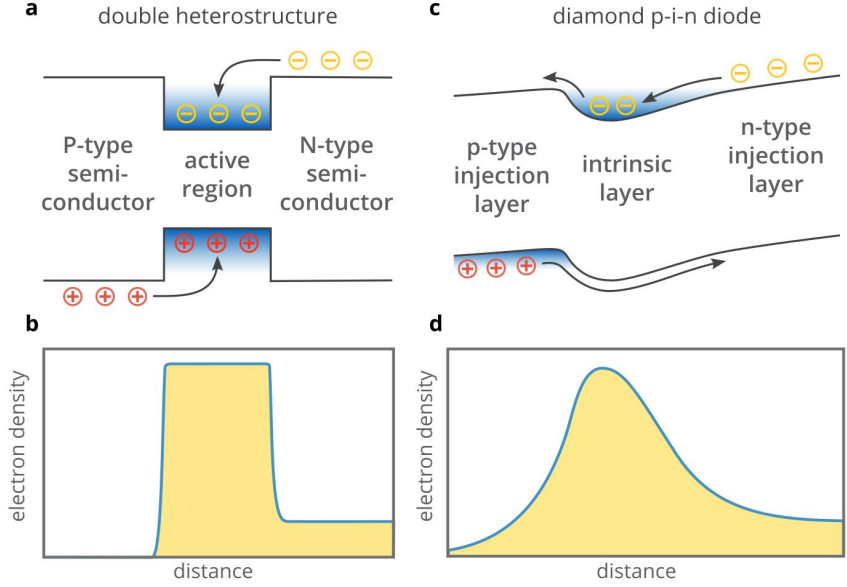

**Figure 1.** (a,b) Schematic illustration of electron and hole injection in a forward biased double heterostructure (a) and qualitative distribution of electron density (b). (c,d) Illustration of electron injection in a diamond p-i-n diode at a high forward bias voltage (c) and spatial distribution of electrons under these conditions (d).

In this work, using a rigorous numerical approach, we investigate for the first time the possibility to observe and exploit the superinjection effect in homojunction diodes based on wide-bandgap semiconductor materials beyond diamond. We discuss in detail how to improve the efficiency of hole injection in SiC p-i-n diodes, which recently demonstrated great promise for quantum applications as single-photon sources [6,23,24]. In addition, we study the superinjection of holes in AlN and ZnS, which are promising materials for optoelectronics [25–27]. In contrast to diamond, these semiconductors do not experience noticeable problems with free electrons but suffer from the high activation energy of acceptors. For example, AlN features an activation energy of acceptors of as high as 0.63 eV [25], i.e., $24k_B T_{300K}$, which limits the density of holes in the p-type doped material to less than $10^{10}$ cm$^{-3}$. However, in this work, we show that at high forward bias voltages, it is possible to significantly overcome this doping limit using the superinjection effect in homojunction p-i-n diodes and inject far more holes into the i-region of the p-i-n diode than the doping of the p-type layer provides.

## Results

### *Superinjection of holes in 4H-SiC diodes*

Figure 2a shows a schematic illustration of the 4H-SiC p-i-n diode. The n-type region of the diode is doped with nitrogen at a concentration of $10^{18}$ cm$^{-3}$. The activation energy of donors is 0.06 eV [28]. The nitrogen donors are partially compensated by acceptor-type defects, the compensation ratio is assumed to be $\eta_n$ = 5%, which provides an electron density of $n_{eqn}$ = 5.5×10$^{17}$ cm$^{-3}$. The p-type injection layer is doped with boron.

The acceptor compensation ratio is also equal to $\eta_p$ = 5% [29]. The concentration of acceptors is $10^{18}$ cm$^{-3}$ and their activation energy equals 0.32 eV [29], which provides a hole density of $p_{eqp}$ = 4.3×10$^{14}$ cm$^{-3}$ in equilibrium. The electron and hole mobilities in the p-, i-, and n-type regions are calculated using the theory from Ref. [30]. The size of the i-region is chosen to be 5 µm, which is high enough to observe the superinjection of holes at moderate currents [14]. Other parameters of the considered 4H-SiC p-i-n diode are listed in Supplementary Information. Using a self-consistent steady-state model, which comprises the Poisson equation, the drift-diffusion current equations and the electron and hole continuity equations to describe the charge carrier behavior in the n-, i- and p-type region of the diode, we performed numerical simulations of the electron and hole transport in the p-i-n diode employing the nextnano++ software (nextnano GmbH, Munich, Germany) and our in-lab-developed simulation tool [8,31,32]. Both methods showed the same results.

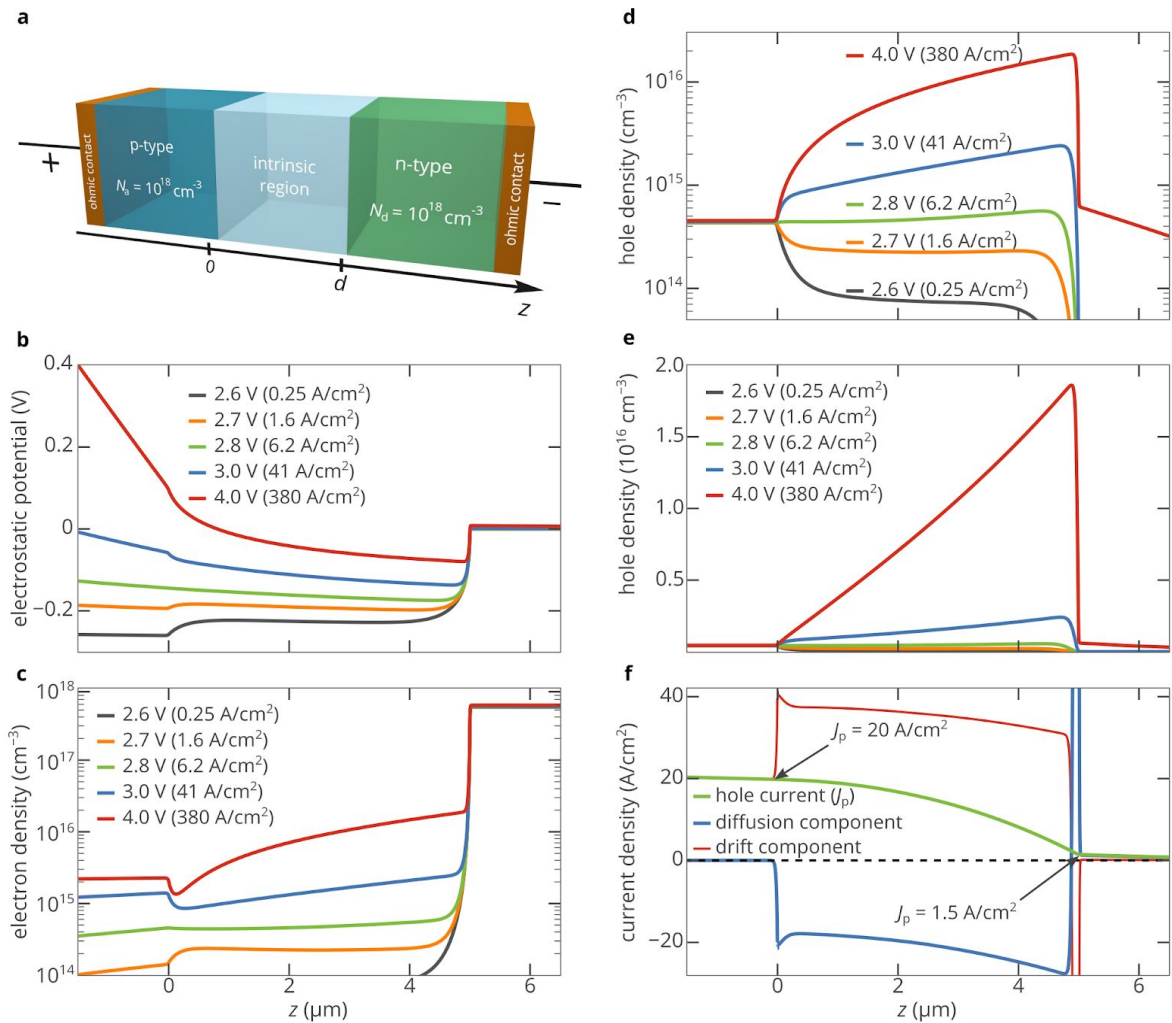

**Figure 2.** (a) Schematic illustration of the 4H-SiC p-i-n diode, $d$ = 5 µm, all parameters of the diode used in the simulations can be found in Supplementary Information (b) Electrostatic potential profile at different bias voltages. (c) Spatial distribution of electrons in the diode at different bias voltages. (d,e) Spatial distribution of holes in logarithmic (c) and linear (d) scales at different bias voltages. (f) Spatial map of the hole current and its components in the diode at $V$ = 4.0 V ($J$=380 A/cm$^2$).

Figure 2 shows the results of the self-consistent numerical simulations of the 4H-SiC p-i-n diode. It is clearly seen that at bias voltages above 2.7 V, a potential well for holes is formed in the i-region of the diode near the i-n junction (Figure 2b). Holes injected from the p-type layer migrate toward the n-type layer and are partially accumulated in this potential well (Figure 2d,e). The higher the bias voltage, the more holes are collected in the well. At the same time, the potential well for holes acts as a potential barrier for electrons. Nevertheless, this barrier is located exactly at the i-n junction (Figure 2b), and electrons easily pass over it via thermionic emission. Moreover, Figure 2c shows that the electron density in the i-region near the i-n junction steadily increases with the current density, which is favorable for light emitting devices, such as electrically pumped single-photon source based on color centers [6,23]. Thus, at current densities above $J$ = 6 A/cm$^2$ ($V$ > 2.8 V), the maximum densities of both electrons and holes in the i-region of the p-i-n diode are found near the i-n junction, which is counterintuitive, since one expects to find the maximum density of electrons in the proximity of the i-n junction and the maximum density of holes in the proximity of the p-i junction, especially in the presence of recombination. Even more surprising is the fact that at a current density of 380 A/cm$^2$, the hole density near the i-n junction exceeds the hole density in the p-type injection layer by a factor of 44 (see Figure 2d,e).

To understand the observed superinjection of holes in the 4H-SiC p-i-n diode, let us look at Figure 2b. It is clearly seen that at high forward bias voltages ($V \gtrsim 3$ V), the band bending is very strong in the p-type region, i.e., the electric field $E$ is high. At the same time, the bands are almost flat, i.e., $E \approx 0$, in n-type region. The reason for this is that the activation energy of acceptors is much higher than the activation energy of donors, and accordingly the density of free carriers in the n-type region is three orders of magnitude higher than in the p-type region. Therefore, at high bias voltages, the drift hole transport dominates in the p-region, while the diffusion hole transport dominates in the n-region, which can be seen in Figure 2f. Thus, in the p-type region near the p-i junction, the hole current can be expressed as

$$J_p|_{z=-0} \approx q\mu_p|_{z=-0} E|_{z=-0} p|_{z=-0}, \qquad (1)$$

where $E$ is the electric field and $\mu_p$ is the hole mobility, and $q$ is the electron charge. On the contrary, in the n-type region near the i-n junction,

$$J_p|_{z=d+0} \approx -qD_p \nabla p \approx qD_p|_{z=d+0} p|_{z=d+0}/L_p, \qquad (2)$$

where $D_p$ is the hole diffusion coefficient, $L_p$ is the hole diffusion length in the n-type layer, and $d$ is the size of the i-region. The hole current at the p-i junction and the hole current at the i-n junction are connected to each other via the current continuity equation:

$$J_p|_{z=d+0} = J_p|_{z=-0} - q\int_0^d R(z)dz, \qquad (3)$$

where $R(z)$ is the net recombination rate. We can rewrite this equation as

$$J_p|_{z=d+0} = J_p|_{z=-0}/K, \qquad (4)$$

where $K$ represents the hole current reduction factor due to recombination in the i-region. If the recombination rate is low, $K \approx 1$ and $J_p|_{z=d+0} \approx J_p|_{z=-0}$. Using equations (1)-(4), we obtain a relation connecting the hole densities at the p-i and i-n junctions:

$$p|_{z=d+0} = \frac{1}{K}\left(\frac{\mu_p|_{z=-0}}{\mu_p|_{z=d+0}} \frac{qL_p}{k_B T}\right) E|_{z=-0} p|_{z=-0}, \tag{5}$$

where $k_B$ is the Boltzmann constant and $T$ is the device temperature.

Equation (5) shows that if the electric field in the p-type layer is strong and the recombination rate in the i-region is low or moderate (i.e., $K$ is not very high), the hole density $p|_{z=d+0}$ in the n-region in the vicinity of the i-n junction can exceed the hole density in the p-type layer. Moreover, the hole diffusion current at the i-n junction is positive, i.e., the gradient of the hole density at $z = d$ is negative. This means that the hole density in the i-region in the vicinity of the i-n junction is even higher than $p|_{z=d+0}$.

Using the material parameters for 4H-SiC listed in Table S1 in Supplementary Information and the hole current reduction factor $K$ obtained in the simulations, we can find that at a bias voltage of 4 V, $p|_{z=d+0}$=7.0x10$^{14}$ cm$^{-3}$, which is 1.6 times higher than the hole density in the p-type injection layer. The obtained value agrees with that obtained in the self-consistent simulations (6.0x10$^{14}$ cm$^{-3}$ (see Figure 2d)). The hole density in the i-region in the vicinity of the i-n junction is even much higher, since the hole diffusion current at the i-n interface, which is proportional to $-\nabla p$, is high and positive. Numerical simulations show that $p$ reaches 1.9x10$^{16}$ cm$^{-3}$ at a distance of 120 nm from the i-n junction (Figure 2d,e), which is 44 times higher than the hole density in the p-type injection layer.

The above explanation shows that the superinjection effect in the 4H-SiC p-i-n diode arises at high bias voltages due the high contrast in the hole transport mechanisms of the p-type and n-type injection layers, which is naturally provided by the very high activation energy of acceptors. The potential well for holes near the i-n junction, which is formed at high bias voltages and where holes are accumulated, is a consequence rather than a cause of the superinjection effect. The real reason for the superinjection effect is the requirement for the current continuity, which can be ensured only by an area with a high hole density between two regions (the p-type and n-type regions of the diode) with different (drift and diffusion) hole transport mechanisms. Therefore, the superinjection of holes can be observed in many wide-bandgap semiconductor p-i-n diodes that feature a similar asymmetry in the hole conduction mechanisms of the n-type and p-type layers. The key requirement for such an asymmetry is a low density of free carriers in the p-type layer and a high density of free carriers in the n-type layer.

Figure 3 shows that the maximum density of non-equilibrium holes created in the i-region in the vicinity of the i-n junction $p_{max}$ increases with the injection current. At a current density of 15 kA/cm$^2$, $p_{max}$ reaches 2.1×10$^{17}$ cm$^{-3}$, which is 500 times higher than the hole density in the p-type injection layer. Although such an injection current seems to be too high for practical applications, a significant improvement of hole injection can be achieved at current densities as low as ~100 A/cm$^2$ (Figure 3).

Finally, we note that the strength of the superinjection effect decreases at extremely high bias voltages (Figure 3), since under these conditions, the drift current becomes significant even in the electron-rich n-type layer. Hence, the diode structure loses the asymmetry in the conductivity mechanisms of the n-type and p-type layer, which is required for the superinjection effect.

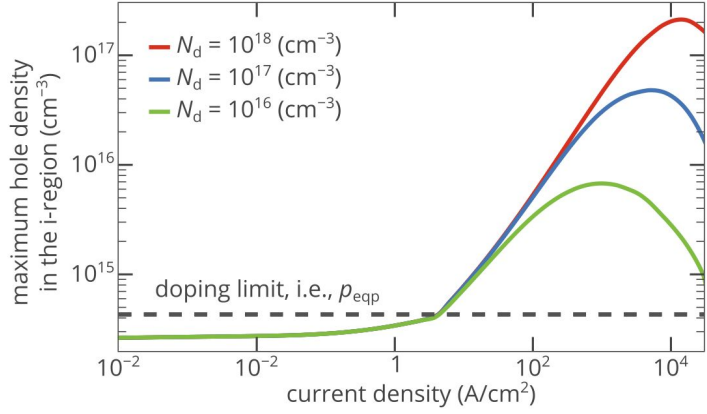

**Figure 3.** Dependence of the maximum hole density in the i-region of the 4H-SiC p-i-n diode on the injection current for different donor concentrations in the n-type injection layer. The 50-nm-thick areas near the p-i and i-n junctions are ignored to avoid overestimation of the density of injected holes.

### *Impact of doping of the n-type injection layer*

As discussed in the previous section, the superinjection effect in the p-i-n diode arises due to the high contrast in the conductivity mechanisms of the p-type and n-type layers. If at a fixed bias voltage, the density of free carriers in the n-type injection layer decreases, the electric field in the n-type layer increases, which directly follows from the Poisson equation. Accordingly, the ratio of the drift hole current to the diffusion hole current increases at the i-n junction. Since the asymmetry in conduction mechanisms of the n-type and p-type injection layers reduces, the strength of the superinjection effect decreases, which can be seen in Figure 3.

### *Impact of the activation energy of acceptors*

The activation energies of donors and acceptors are critical parameters that can strongly affect the efficiency of hole injection. Since the activation energy of donors in 4H-SiC is much lower than the activation energy of acceptors [29], we focus only on the latter. The density of holes in the p-type layer rapidly decreases as the activation energy of acceptors $E_A$ increases, especially at a non-zero acceptor compensation ratio [33]:

$$p_{eqp} = \frac{\eta_p N_a + p_1}{2}\left(\sqrt{1 + \frac{4 N_a (1-\eta_p) p_1}{(\eta_p N_a + p_1)^2}} - 1\right), \quad (6)$$

where

$$p_1 = \frac{1}{g_A} N_v \exp\left(-\frac{E_A}{k_B T}\right), \quad (7)$$

Here, $N_a$ is the acceptor concentration, $g_A$ is the degeneracy factor of the acceptor level, and $N_v$ is the effective density of states in the valence band. Figure 4a shows the dependence of the hole density on the activation energy of acceptors for an acceptor compensation ratio of $\eta_p$=5%. At $E_A$ = 0, the density of holes in the p-type region $p_{eqp}$ is as high 8.2×10$^{17}$ cm$^{-3}$. However, at $E_A$ = 0.23 eV, $p_{eqp}$ decreases by a factor of 70, to 1.14×10$^{16}$ cm$^{-3}$. This activation energy corresponds to aluminum, which is the most shallow acceptor in 4H-SiC [29]. At $E_A = 0.5$ eV, $p_{eqp}$ is as low as $4.1\times10^{11}$ cm$^{-3}$, i.e., six orders of magnitude lower than the concentration of acceptors. However, the lower the density of holes in the p-type layer, the higher the asymmetry in the conduction properties of the p-type and n-type regions, and therefore the stronger the superinjection effect in the p-i-n diode.

Figure 4b shows that the superinjection effect is very weak and can hardly be detected if the activation energy of acceptors is lower than 0.15 eV. If 4H-SiC is doped with aluminum ($E_A = 0.23$ eV), the maximum density of injected holes can exceed the hole density in the p-type layer more than tenfold (see Figure 4b). As the activation energy of acceptors further increases, the superinjection effect strengthens (Figure 4b-d). Moreover, it can be achieved at lower currents (Figure 4b). Figure 5 shows that although the activation energy of boron (0.32 eV) is significantly higher than the activation energy of aluminum (0.23 eV), at room temperature and current densities above 500 A/cm$^2$, the superinjection effect allows the boron doped p-type layer to inject roughly the same density of holes as the aluminum doped layer can inject at the same current density. At $E_A > 0.4$ eV, a greatly improved hole injection can be obtained at current densities below 10 A/cm$^2$, which is particularly interesting for practical applications. At higher injection currents, the strength of the superinjection effect can exceed 1000. However, we should note that such a very high improvement of the hole injection efficiency is achieved only in a relatively narrow region near the i-n junction (see Figure 4d and red curve in Figure 4c). Nevertheless, this feature could be advantageous for some applications, such as single-photon sources based on color centers [6,23].

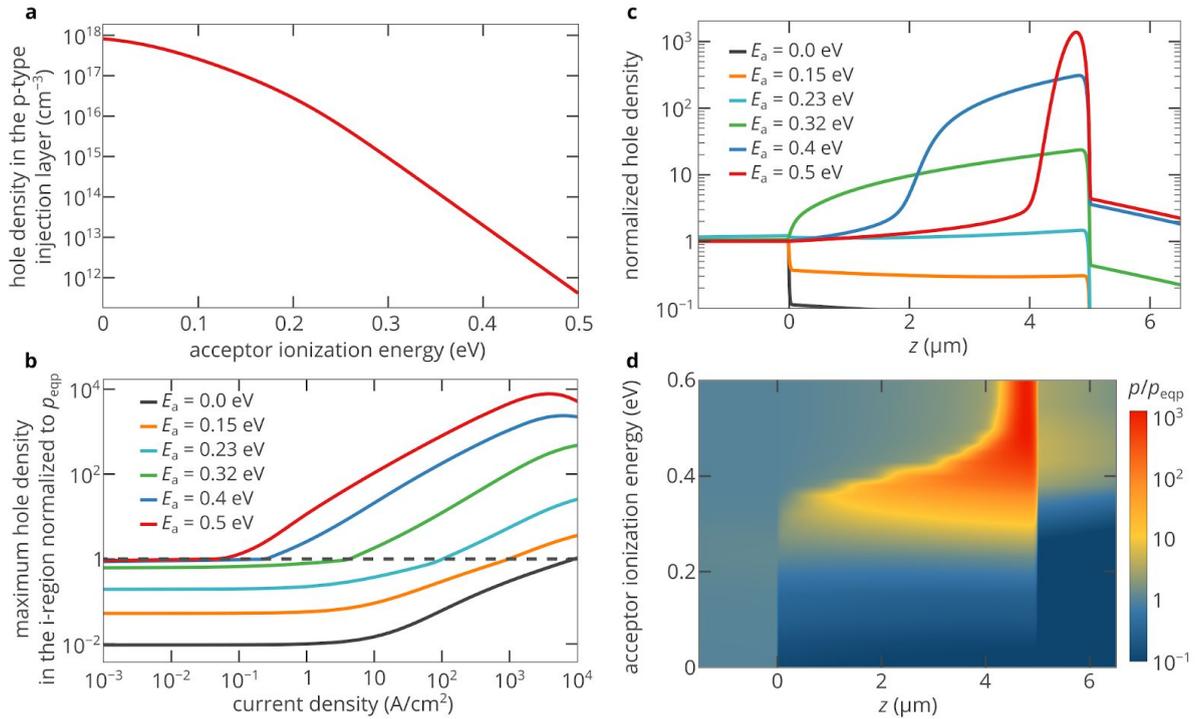

**Figure 4.** (a) Dependence of the hole density in the p-type injection layer on the acceptor ionization energy. The acceptor compensation ratio is equal to 5%. (b) Dependence of the maximum hole density in the i-region of the 4H-SiC p-i-n diode, which is normalized to the density of holes in the p-type layer $p_{eqp}$, on the injection

current for different activation energies of acceptors in the p-type injection layer. The 50-nm-thick areas near the p-i and i-n junctions are ignored to avoid overestimation of the density of injected holes. (c) Spatial distribution of holes in the 4H-SiC p-i-n diode for different activation energies of acceptors in the p-type injection layer at an injection current density of 200 A/cm$^2$. The hole density is normalized to the density of holes in the p-type injection layer. (d) Variation of the spatial distribution of holes in the 4H-SiC p-i-n diode with the activation energy of acceptors in the p-type injection layer at an injection current density of 200 A/cm$^2$. The hole density is normalized to the density of holes in the p-type injection layer.

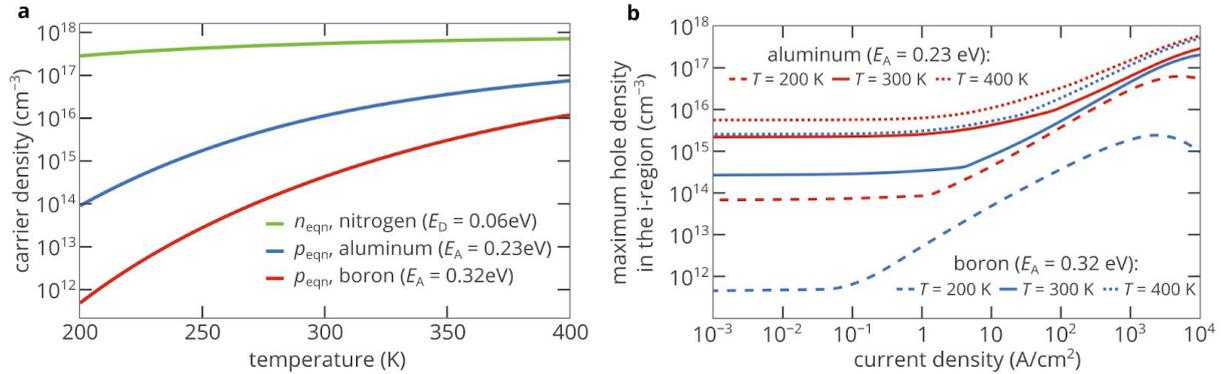

**Figure 5.** (a) Electron density in the n-type layer and hole density in the p-type layer as functions of temperature. (b) Dependence of the maximum hole density in the i-region of the 4H-SiC p-i-n diode on the injection current for three different temperatures and two different activation energies of acceptors. The 50-nm-thick areas near the p-i and i-n junctions are ignored to avoid overestimation of the density of injected holes. The material parameters of 4H-SiC at 200 K and 400 K are provided in Supplementary Information.

## *Impact of temperature*

Equations (6) and (7) show that the densities of free carriers in the p-type and n-type injection layers greatly depend on the device temperature. Since the activation energy of acceptors is very high, even a small decrease in temperature significantly decreases the hole density. If the p-type layer of the diode is doped with aluminum, the density of holes decreases by a factor of 160 as the temperature drops from 300 K to 200 K (Figure 5a). In the boron doped p-type layer, the reduction in the hole density is even stronger: $p_{eqp}$ drops by a factor of 880. However, this decrease in the free carrier density increases the asymmetry in the conduction properties of the n-type and p-type injection layer, which is beneficial for the superinjection effect, as discussed in the previous section. Figure 5b shows that although at $T$ = 200 K, $p_{eqp}$ is 160 lower than $T$ = 300 K, the density of injected holes is only twice lower than at room temperature at current densities above 300 A/cm$^2$. On the contrary, at high temperatures, the density of holes in the p-type layer is higher, and the superinjection effect is weaker, i.e., the ratio of the maximum injected hole density to the density of holes in the p-type layer is significantly lower than at room temperature.

### 6H-SiC and 3C-SiC p-i-n diodes

Activation energies of acceptors and other material parameters in 6H-SiC are roughly the same as in 4H-SiC [29]. Accordingly, the strength of the superinjection effect and its other properties are very similar to that in 4H-SiC p-i-n diodes. Since 4H-SiC p-i-n diodes are already discussed, in this section, we focus only on 3C-SiC.

The activation energies of acceptors in 3C-SiC are pretty much the same as in 4H-SiC: 0.26 eV for Al [34] and 0.34 eV for Ga [34]. The electron and hole mobilities are slightly lower than in 4H-SiC [35,36] and the carrier lifetime is slightly longer [37,38]. Equation (5) shows that the superinjection effect is not very sensitive to the carrier mobility, but inversely proportional to the carrier lifetime. Therefore, the hole injection in the 3C-SiC p-i-n diode is expected to be better than in the 4H-SiC diode considered above, which agrees with the results of the self-consistent numerical simulations of the 3C-SiC p-i-n diode (see Figure 6).

### ZnS and AlN diodes

Although the activation energies of acceptors in silicon carbide are high, which limits the density of holes in the p-type material to $10^{14}$ - $10^{15}$ cm$^{-3}$, the p-type doping problem is even more pronounced in such materials as gallium nitride, aluminum nitride or zinc sulfide. If in the case of GaN (which also features very high activation energies of acceptors [39]), one can use AlGaN/GaN/AlGaN heterostructures [40], the extremely high bandgap energy of AlN of 6.2 eV simply excludes this option. At the same time, the activation energy of the most shallow acceptor in AlN is about 0.6 eV, which limits the density of holes in the p-type material to ~$10^{10}$ cm$^{-3}$ [25]. Therefore, it is crucially important to improve the hole injection efficiency in AlN homojunction electronic and optoelectronic device. Figure 6b and the blue curve in Figure 6d show the results of the numerical simulations for AlN p-i-n diode. For the parameter of AlN, see Table S1 in Supplementary Information. Despite that the activation energy of acceptors in AlN is much higher than in silicon carbide, the maximum strength of the superinjection effect is lower, which is explained by lower mobility, slightly lower carrier lifetime and lower density of electrons in the n-type layer due to the higher activation energy of donors in AlN (0.25 eV [41]). The latter factor is especially important since the density of electrons in the n-type layer limits the maximum voltage at which equations (1) and (2) are valid and therefore the maximum value of the electric field in equation (5). In addition, the spatial distribution of the density of injected holes is remarkably different (Figure 6b). Due to the low density of free carriers in the i-region of the diode even at high bias voltages, the electric field in the i-region is strong. The electrostatic potential at the p-i junction is well above the electrostatic potential at the i-n junction. Hence, the potential well in the i-region is narrow, and a significant improvement in the density of injected holes is achieved only in a relatively narrow region near the i-n junction. Nevertheless, owing to the superinjection effect, at high bias voltages, the density of injected holes can exceed the doping limit by two orders of magnitude.

Another example of a wide-bandgap semiconductor, which can hardly be used as an active layer of the double heterostructure is zinc sulfide [26]. At the same time, ZnS has long been known as a phosphor material owing to the efficient luminescence of different point defects [42]. However, electrical excitation of these defects was always a problem due to the extremely high activation energy of acceptors (~0.6 eV [43]), which limits the density of holes in the p-type layer to less than $10^{10}$ cm$^{-3}$. Therefore, it was proposed not to use p-n and p-i-n structures and try to implement the hole injection mechanism in metal/insulator/semiconductor structures [44]. Figure 6b and the red curve in Figure 6d show the simulated efficiency of hole injection in the ZnS p-i-n diode. For the parameters of the diode, see Table S1 in Supplementary Information. These figures show that the hole density in the i-region of the structure is lower than $6\times10^9$ cm$^{-3}$ until the current density exceeds 0.02 A/cm$^2$. Above 0.02 A/cm$^2$, the maximum density of injected holes rapidly increases with the injection current and can reach $7.8\times10^{12}$ cm$^{-3}$, which is three orders of magnitude higher than the hole density in the p-type layer. Figure 6c shows that the superinjection effect in ZnS is very similar to that in AlN. However, its maximum strength is higher (mostly due to the higher density of electrons in the n-type layer), and the effect can be observed at lower currents due to the longer carrier lifetimes (see Table S1 in Supplementary Information).

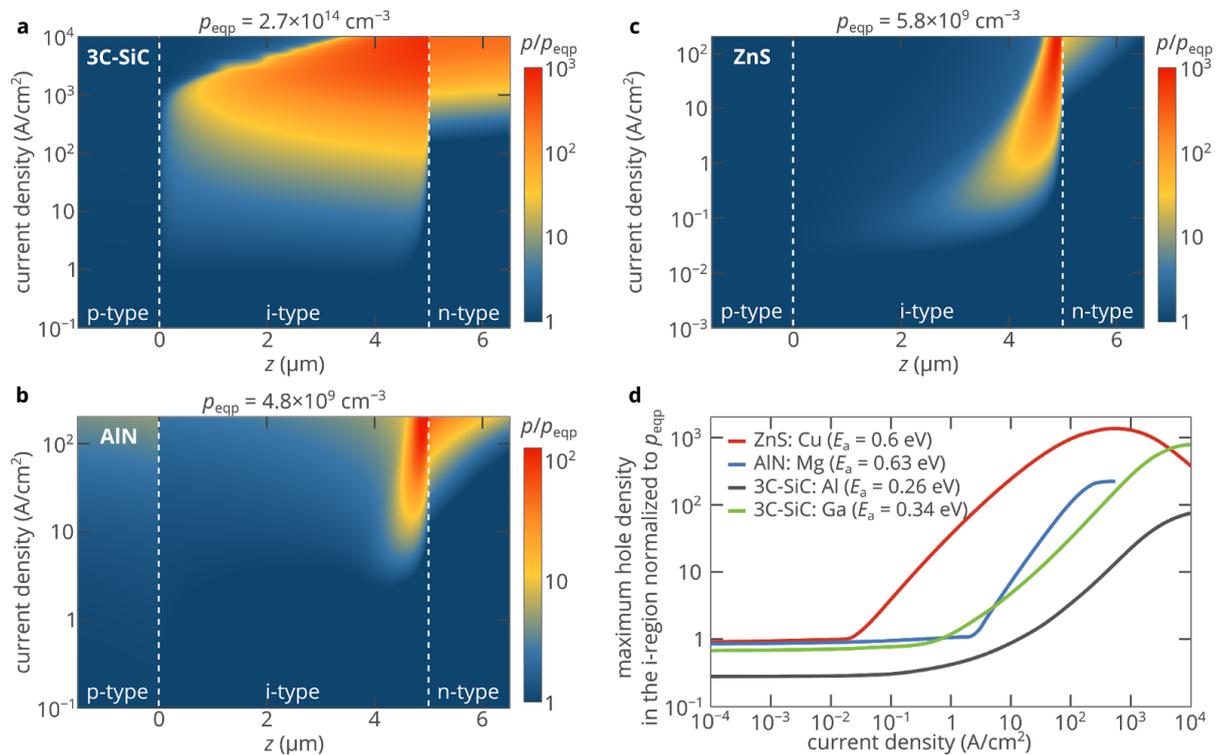

**Figure 6.** (a-c) Spatial distribution of the hole density in the i-region of the three p-i-n diodes based on 3C-SiC with the p-layer doped with gallium (a), AlN (b), and ZnS (c). The hole density is normalized to the hole density in the p-type layer. (d) Dependence of the maximum hole density in the i-region of different the p-i-n diodes. The 50-nm-thick areas near the p-i and i-n junctions are ignored to avoid overestimation of the density of injected holes. The hole density is normalized to the density of holes in the p-type injection layer. For the 3C-SiC diode, two dopants (Al and Ga) of the p-type layer are considered. The parameters of the 3C-SiC, ZnS, and AlN diodes are listed in Supplementary Information.

**Discussion**

We have numerically demonstrated the superinjection of holes in homojunction p-i-n diodes based on different wide-bandgap semiconductors. This effect gives the possibility to create a high density of non-equilibrium holes in the i-region of the p-i-n diode at high forward bias voltages. The created density of holes exceeds the hole density in the p-type injection layer by more than two orders of magnitude, which can be exploited to improve the performance of light emitting devices. The highest hole density is found near the i-n junction, which contradicts the common belief about p-i-n diodes, since in the i-region, one could expect to find the highest hole density near the p-i junction. However, we have provided a clear physical explanation for the observed effect, which is based on the asymmetry in the free carrier densities in the n-type and p-type layers. In SiC, GaN, AlN, ZnS and many wide-bandgap semiconductors, this asymmetry is naturally provided by the high activation energy of acceptors, which greatly limits the density of holes in the p-type layer of the diode. Hence, for optoelectronic devices, it is crucially important to overcome this doping limit, especially if it is not possible to use heterostructures. We have demonstrated that the superinjection of holes can be observed at high forward bias voltages. At the same time, the superinjection effect does not show itself at low and moderate injection currents (bias voltages). The higher the density of holes in the p-type layer, the higher the threshold current (see Figure 4b). This is one of the reasons, why the effect was not reported previously.

The strength of the superinjection effect depends on the minority carrier lifetime and other properties of the semiconductor diode, but the most crucial parameter is the activation energy of acceptors $E_A$ since it determines the hole density in the p-type region. We found that the superinjection of holes can hardly be detected at doping levels above $10^{17}$ cm$^{-3}$ if $E_A$ is lower than ~0.15 eV. On the other hand, at high activation energies ($E_A \gtrsim 0.4$ eV), it is strong even at relatively low injection currents. Assuming the density of electrons to be high in the n-type region of the p-i-n diode ($n_{eqn} \gtrsim 10^{17}$ cm$^{-3}$), the results of the numerical simulations and equation (5) show that the higher the activation energy of acceptors, the stronger the effect is. However, we should emphasize that the absolute value of the density of injected holes in a diode with a high activation energy of acceptors in the p-type layer (e.g., $E_A = 0.32$ eV) cannot exceed the absolute value of the density of injected holes in a diode with a low activation energy of acceptors in the p-type layer (e.g., $E_A = 0.23$ eV) (see Figure 5). Therefore, the key to the improvement of the hole injection efficiency is a low activation energy of acceptors. However, if $E_A$ cannot be reduced, the superinjection effect is the only possibility to improve the hole injection efficiency in homojunction diodes.

A disadvantage of the superinjection effect is that when a very high density of injected holes is achieved, the electron current is at least an order of magnitude higher than the hole current. Therefore, the energy efficiency of LEDs based on the superinjection effect in homojunction p-i-n diodes will be lower than that of based on double heterostructures. A strong effect also requires high bias voltages (>6 V for 4H-SiC diodes and >30 V for AlN) that can create a high electric field in the p-type region of the diode (see equation (5)). Accordingly, the heat generation rate is considerably high:

more than 100 W/cm$^2$ for the ZnS diode and more than ~1 kW/cm$^2$ for other considered diodes. This should also be taken into account in practical applications.

We should emphasize that the superinjection effect not only gives a unique possibility to create a density of non-equilibrium holes which is orders of magnitude higher than the doping limit $p_{eqp}$, but also provides it in the vicinity of the i-n junction, where the density of injected electrons is maximal (see Figure 2c,d). Thus, we obtain the maximum electron and hole densities at roughly the same place. This feature is extremely important [23] for the design and development of ultrabright electrically driven single-photon sources based on point defects in SiC, AlN, ZnO and other wide-band gap semiconductors [6,23,24,27,45–47], which are key elements for many applications of quantum information science, such as quantum-communication networks and optical quantum computers [48,49]. We believe that our findings will stimulate research focused on the development of novel high-performance light source based on wide-bandgap semiconductors.

## Author Contributions

I.A.K. performed the numerical simulations. D.Y.F. conceived the idea and wrote the manuscript. Both authors analyzed the results.

## Funding

The work is supported by the Russian Science Foundation (17-79-20421).

## Conflicts of Interest

The authors declare no conflict of interest.

# Supplementary Materials for
# Superinjection of holes in homojunction diodes based on wide-bandgap semiconductors

Igor A. Khramtsov and Dmitry Yu. Fedyanin*

*Laboratory of Nanooptics and Plasmonics, Moscow Institute of Physics and Technology, 141700 Dolgoprudny, Russian Federation*

*E-mail: dmitry.fedyanin@phystech.edu

**Table S1.** Main material parameters used in the numerical simulations.

|  | **4H-SiC** | **3C-SiC** | **wz-ZnS** | **wz-AlN** |
|---|---|---|---|---|
| **Energy bandgap at $T$=300 K, eV** | 3.23 [1] | 2.36 [1] | 3.8 [2] | 6.23 [3] |
| **Dielectric constant** | 9.78 [4] | 9.72 [5] | 8.32 [2] | 9.14 [6] |
| **Density of states effective electron mass** | $0.77m_0$ [7] | $0.72m_0$ [8] | $0.31m_0$ [2] | $0.31m_0$ [3] |
| **Density of states effective hole mass** | $0.91m_0$ [9] | $1.11m_0$ [10] | $0.7m_0$ [11] | $7.26m_0$ [12] |
| **Acceptor concentration in the p-type region, cm$^{-3}$** | $10^{18}$ | $10^{18}$ | $10^{18}$ | $10^{18}$ |
| **Acceptor compensation ratio** | 5% | 5% | 5% | 40% (fitted from the experimental data) [13] |
| **Acceptor ionization energy, eV** | 0.23 (Al) [14] 0.32 (B) [15] | 0.26 (Al) [16] 0.34 (Ga) [16] | 0.6 (Cu) [17] | 0.63 (Mg) [13] |
| **Donor concentration in the n-type region, cm$^{-3}$** | $10^{18}$ | $10^{18}$ | $10^{18}$ | $10^{18}$ |

| **Donor compensation ratio** | 5% | 5% | 5% | 10 % [18] |
|---|---|---|---|---|
| **Donor ionization energy, eV** | 0.06 (N) [19] | 0.05 (N) [16] | 0.1 (Al) [20] | 0.25 (Si) [21] |
| **Electron mobility in the p-type region, cm$^2$/Vs** | 500 [22,23] | 200 [24] | 100 [2] | 100 [21] |
| **Hole mobility in the p-type region, cm$^2$/Vs** | 140 [22,23] | 55 [25] | 10 [26] | 5 [25] |
| **Electron mobility in the i-type region, cm$^2$/Vs** | 900 [22,23] | 650 [24] | 140 [2] | 300 [25] |
| **Hole mobility in the i-type region, cm$^2$/Vs** | 140 [22,23] | 70 [25] | 15 [26] | 14 [25] |
| **Electron mobility in the n-type region, cm$^2$/Vs** | 300 [22,23] | 200 [24] | 100 [2] | 100 [21] |
| **Hole mobility in the n-type region, cm$^2$/Vs** | 120 [22,23] | 55 [25] | 10 [26] | 5 [25] |
| **Electron saturation velocity, cm/s** | 2.2×10$^7$ [27] | 2.0×10$^7$ [1] | 1.8×10$^7$ (ZnSe value) [28] | 1.5×10$^7$ [25] |
| **Hole saturation velocity, cm/s** | 1.5×10$^7$ [1] | 1.5×10$^7$ [1] | 1.8×10$^7$ (the same as for electrons) | 1.25×10$^7$ [25] |
| **SRH lifetime, ns** | 20 [29] | 100 [30] | 100 (estimated from Refs. [31–33]) | 10 (GaN value) [34,35] |
| **Radiative recombination coefficient, cm$^3$/s** | - | - | 7.1×10$^{-10}$ (ZnTe value) [36] | 0.45×10$^{-10}$ [37] |

**Table S2.** Temperature dependence of the main material parameters of 4H-SiC.

| | $T = 200$ K | $T = 400$ K |
|---|---|---|
| **Energy bandgap, eV** | colspan: $E_g(T) = E_g(0) - 6.5\times10^{-4}T^2/(1300+T)$ [1] | |
| **Electron mobility in the p-type region, cm$^2$/Vs** | 730 [22,23] | 260 [22,23] |
| **Hole mobility in the p-type region, cm$^2$/Vs** | 450 [22,23] | 70 [22,23] |
| **Electron mobility in the i-type region, cm$^2$/Vs** | 1550 [22,23] | 310 [22,23] |
| **Hole mobility in the i-type region, cm$^2$/Vs** | 345 [22,23] | 75 [22,23] |
| **Electron mobility in the n-type region, cm$^2$/Vs** | 400 [22,23] | 200 [22,23] |
| **Hole mobility in the n-type region, cm$^2$/Vs** | 245 [22,23] | 70 [22,23] |
| **SRH recombination lifetime, ns** | colspan: $\tau_{SRH} \sim (T/300)^2$ [38,39] | |
| **Electron saturation velocity, cm/s** | colspan: $V_{satn} \sim (T/300)^{-0.44}$ [27,40,41] | |
| **Hole saturation velocity, cm/s** | colspan: The same temperature dependence as for electrons | |